\documentclass{WileyMSP-template}
\usepackage{graphicx}
\usepackage{mathtools}
\usepackage{amsmath}
\usepackage{amssymb}
\usepackage{bbm}
\usepackage{bm}
\usepackage{cancel}
\usepackage{xspace}
\usepackage{braket}
\usepackage{xcolor}
\usepackage{siunitx}
\usepackage[english]{babel}
\usepackage{calrsfs}
\usepackage{comment}
\usepackage{placeins}
\usepackage[export]{adjustbox}
\usepackage{wrapfig}
\usepackage[superscript]{cite}
\usepackage[colorlinks=true]{hyperref} 
 
\newcommand{\ee}{\mathrm{e}}  
\DeclareMathOperator*{\ii}{i} 
\newcommand*\dd{\mathop{}\!\mathrm{d}}

\renewcommand{\vec}[1]{\bm{#1}} 
\newcommand{\mat}[1]{\bm{#1}} 
\newcommand{\kel}[1]{\underline{#1}} 

\definecolor{hblue}{RGB}{0,150,255}

\newcommand{\resub}[1]{{\color{black} #1}}

\begin{document}

\pagestyle{fancy}
\setlength{\headheight}{24.81955pt}
\rhead{\includegraphics[width=2.5cm]{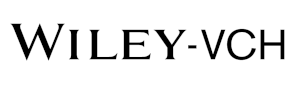}}

\title{Iterated Perturbation Theory for Mott Insulators in a Static Electric Field with Optical Phonons}

\maketitle

\author{Tommaso Maria Mazzocchi*}
\author{Enrico Arrigoni*}

\dedication{}

\begin{affiliations}
T. M. Mazzocchi, E. Arrigoni \\
Institute of Theoretical and Computational Physics, Graz University of Technology, Petersgasse 16/II, 8010 Graz (Austria) \\
mazzocchi@tugraz.at, arrigoni@tugraz.at

\end{affiliations}


\begin{abstract}

This manuscript aims to compare the so-called iterated perturbation theory (IPT) and auxiliary master equation approach (AMEA) impurity solvers for a Mott insulating system driven out of equilibrium by a static electric field. Electronic heat bath and optical phonons are employed as dissipation mechanism of the current-induced Joule heat that the excited electrons of the lattice experience as the result of the field's driving. \resub{Despite its simplicity, the IPT approach yields results which qualitatively are in good agreement with those obtained within the AMEA impurity solver, although fails to reproduce some correlation effects.} 

\end{abstract}

\section{Introduction}

The last decade has witnessed an increasing interest in the physics of Mott insulating systems as they can undergo phase transitions when driven out of equilibrium by both statically and periodically electric fields~\cite{aron.12,am.we.12,ar.ko.12,li.ar.15,ha.li.18,mu.we.18}. Due to this property, they could be used to model the so-called {\em insulator-to-metal transition}, which has been investigated theoretically in the seminal works~\cite{le.ra.85,be.ki.94} and then observed experimentally~\cite{st.ca.13,ja.tr.15}. Due to their strongly interacting nature, Mott insulating systems require a nonperturbative method to be dealt with. Nowadays, the most well-established approach is given by the dynamical mean-field theory (DMFT)~\cite{me.vo.89,ge.ko.92,ge.ko.96,ko.sa.06,fr.tu.06,ao.ts.14} which \resub{holds} under both equilibrium and non-equilibrium conditions. The DMFT relies on impurity solvers to address the non-equilibrium steady-state of the system, which often are the bottleneck of the approach as they can be computationally costly. Purpose of this manuscript is to benchmark the results of the iterated perturbation theory (IPT) impurity solver, which is known to work well in both the very weak and very strong interacting {\em regime}~\cite{ka.ko.96}, against those obtained within the so-called auxiliary master equation approach (AMEA) impurity solver~\cite{ar.kn.13,do.nu.14,do.ga.15,ti.do.15,we.lo.23}. The IPT solver yields results which are in quite good agreement with those obtained by employing AMEA for the very same setup investigated in a previous work from the authors~\cite{ma.we.23}. The rest of the manuscript is organized as follows: In Section~\ref{sec:MO_HA} we introduced the model at hand while Section~\ref{sec:methods} will be devoted to a short recap of the Green's function formalism and the IPT impurity solver. Results are discussed in Section~\ref{sec:results} while Section~\ref{sec:conclusion} is left for conclusions and further considerations. The details of the physical setup under investigation and the derivation of all the relevant observables of interest can be found in the previous work from the authors~\cite{ma.we.23}.

\section{Model Hamiltonian}\label{sec:MO_HA}

We study the single-band Hubbard model in the presence of a constant electric field in the temporal gauge~\cite{ma.we.23}, the Hamiltonian of which is given by
\begin{equation}\label{eq:MicroHamiltonian}
\hat{H}(t) = \hat{H}_{\text{\tiny U}}(t) + \hat{H}_{\text{bath}} + \hat{H}_{\text{e-ph}} + \hat{H}_{\text{ph}}.
\end{equation}
\resub{The units employed in this manuscript are such that the lattice spacing $a$, Planck constant $\hbar$ and the electron charge $q$ are chosen as $a=\hbar=-q=1$}, hence the Hubbard Hamiltonian $\hat{H}_{\text{\tiny U}}(t)$ in Equation~\eqref{eq:MicroHamiltonian} is given by
\begin{equation}\label{eq:Hubbard_ham}
\hat{H}_{\text{\tiny U}}(t) = \varepsilon_{\text{c}} \sum_{i\sigma}\hat{n}^{f}_{i\sigma} -\sum_{\sigma}\sum_{(i,j)} \underbrace{t_{\text{c}} \ \ee^{-\ii \left( \vec{r}_j - \vec{r}_i \right) \cdot \vec{A}(t)}}_{= \ t_{ij}(t)} \hat{f}^{\dagger}_{i\sigma} \hat{f}_{j\sigma} + U \sum_{i} \hat{n}^{f}_{i\uparrow} \hat{n}^{f}_{i\downarrow},
\end{equation}
where $\hat{f}^{\dagger}_{i\sigma}$ ($\hat{f}_{i\sigma}$) is the creation (annihilation) operator of an electron of spin $\sigma= \{ \uparrow,\downarrow \}$ at the $i$-th lattice site and $\hat{n}^{f}_{i\sigma}\equiv \hat{f}^{\dagger}_{i\sigma} \hat{f}_{i\sigma}$ the corresponding density operator. Sums over nearest neighbor sites are denoted by $(i,j)$. The electrons' \emph{onsite energy} is chosen as $\varepsilon_{\text{c}} \equiv -U/2$ for the system to be fulfil particle-hole symmetry and $t_{\text{c}}$ is the {\em bare} hopping amplitude. The homogeneous vector potential $\vec{A}(t)= -\vec{F}\  t$ is chosen such that the static electric field is constant and oriented along the body diagonal of a {\em hypercubic} lattice $\vec{e}_{0}=(1,1,\ldots,1)$ and is given by $\vec{F}= -\partial_{t}\vec{A}(t)$.

We take the infinite-dimension limit,~\cite{tu.fr.06,mu.we.18} i.e. $d \rightarrow \infty$, with the usual rescaling of the hopping $t_{\text{c}}=t^{\ast}/(2\sqrt{d})$, which allows to perform summations over the electron crystal momentum using the joint density of states~\cite{ts.ok.08,ma.ga.22} $\rho(\epsilon,\overline{\epsilon}) = 1/(\pi t^{\ast 2}) \ \exp[-( \epsilon^{2} + \overline{\epsilon}^{2})/t^{\ast 2}]$ with $\epsilon = -2t_{\text{c}} \sum_{i=1}^{d} \cos k_i$ and $\overline{\epsilon} = -2t_{\text{c}}\sum_{i=1}^{d} \sin k_i$.

An optical phonon branch is {\em attached} to each lattice site by means of the Hamiltonian
\begin{equation}\label{eq:e-ph_Ein_ham}
\hat{H}_{\text{e-ph}} = g \sum_{i\sigma} \hat{n}^{f}_{i\sigma} \hat{x}_{i}
\end{equation}
with $\hat{x}_{i}\equiv (\hat{b}^{\dagger}_{i} + \hat{b}_{i})/\sqrt{2}$, where $\hat{b}^{\dagger}_{i}$ ($\hat{b}_{i}$) creates (annihilates) an optical phonon with energy $\omega_{\text{E}}$ at the lattice site $i$. The optical phonon Hamiltonian consists of an Einstein phonon $\hat{H}_{\text{ph},\text{E}} = \omega_{\text{E}}\sum_{i}\hat{n}^{b}_{i}$ with $\hat{n}^{b}_{i}=\hat{b}^{\dagger}_{i}\hat{b}_{i}$ the phonon density, coupled to a noninteracting, ohmic bath $\hat{H}_{\text{ph},\text{ohm}}$ with spectral density given in Equation~\eqref{eq:ohm_bath_spec}.

To stabilize the DMFT loop we include electronic heat baths, consisting of a collection of noninteracting fermionic degrees of freedom, coupled {\em locally} to each lattice site which are described by the Hamiltonian $\hat{H}_{\text{bath}}$, the details of which will be specified in Section~\ref{sec:Dyson-eq}, see Equation~\eqref{eq:ret_sigma_bath}.

\section{Methods}\label{sec:methods}

\subsection{Green's function formalism}

This section contains a short recap of the Green's function (GF) approach, a versatile tool for the solution of many-body problems in- and out-of-equilibrium~\cite{schw.61,keld.65,ra.sm.86,ha.ja,st.va.13} which has been successfully applied to several systems in condensed matter over the last decade~\cite{aron.12,am.we.12,ar.ko.12,ec.we.13.db,li.ar.15,ha.li.18,di.ha.23,ha.ar.23}. 

\subsubsection{Electron Dyson equation}\label{sec:Dyson-eq}
The interacting electron GF obeys the Dyson equation 
\begin{equation}\label{eq:FullDysonEq}
\kel{\mat{G}}^{-1}(\omega,\epsilon,\overline{\epsilon}) = \kel{\mat{G}}^{-1}_{0}(\omega,\epsilon,\overline{\epsilon}) - \kel{\mat{\Sigma}}_{\text{bath}}(\omega) - \kel{\mat{\Sigma}}(\omega) - \kel{\mat{\Sigma}}_{\text{e-ph}}(\omega)
\end{equation}
with $\kel{\mat{G}}_{0}$ denoting the GF corresponding to the noninteracting part of the Hamiltonian in Equation~\eqref{eq:Hubbard_ham}. By $\kel{\vec{\Sigma}}_{\text{bath}}$, $\kel{\mat{\Sigma}}$ and $\kel{\mat{\Sigma}}_{\text{e-ph}}$ we denote the fermionic heat bath, electron and electron-phonon (e-ph) self-energy (SE), respectively. By means of the DMFT and Migdal~\cite{mu.we.15,mu.ts.17} approximations both $\kel{\mat{\Sigma}}$ and $\kel{\mat{\Sigma}}_{\text{e-ph}}$ in Equation~\eqref{eq:FullDysonEq} are {\em local}.

Every quantity $\kel{\vec{X}}$ with an underline denotes the so-called {\em Keldysh structure}, namely
\begin{equation}\label{eq:Keld-structure}
\kel{\mat{X}} \equiv 
\begin{pmatrix}
\mat{X}^{\text{R}} & \mat{X}^{\text{K}}\\
\mat{0}         & \mat{X}^{\text{A}} \\
\end{pmatrix}
\end{equation}
with $\mat{X}^{\text{R},\text{A},\text{K}}$ being the {\em retarded}, {\em advanced} and {\em Keldysh} components where $\mat{X}^{\text{K}}\equiv \mat{X}^{>} + \mat{X}^{<}$ and $\mat{X}^{\gtrless}$ being the {\em greater} and {\em lesser} components. Each of the Keldysh components in Equation~\eqref{eq:Keld-structure} is then a matrix in the {\em Floquet indices}~\cite{jo.fr.08,ts.ok.08,so.do.18,ma.ga.22,ga.ma.22,ma.we.23}. However, given the time-translation invariant character of the problem at hand,~\cite{ts.ok.08,so.do.18,ma.ga.22,ma.we.23} only the diagonal components and especially the {\em time-averaged} element $\kel{\vec{X}}_{00}$ are non-vanishing~\footnote{We recall that due to the fundamental property of Floquet-represented matrices, i.e. $\kel{\vec{X}}_{mn}(\omega) = \kel{\vec{X}}_{m-n,0}(\omega+nF)$, $F$ being the applied field, the diagonal entries $\kel{\vec{X}}_{mn}$ can be obtained by the $\kel{\vec{X}}_{00}$ alone.}. For this reason we will omit the Floquet indices in the rest of the manuscript.

Details concerning the computation of the e-ph SE $\kel{\vec{\Sigma}}_{\text{e-ph}}$ within the Migdal approximation can be found in our previous work~\cite{ma.ga.22,ma.we.23}.

We recall the definition of the electron spectral function (SF)
\begin{equation}
 A(\omega)\equiv -\frac{1}{\pi}\text{Im}\left[ G^{\text{R}}_{\text{loc}}(\omega) \right],
\end{equation}
where the {\em local} electron GF is given by
\begin{equation}\label{eq:local_GF}
 G^{\text{R}}_{\text{loc}}(\omega) = \int \dd \epsilon \int \dd \overline{\epsilon} \ \rho(\epsilon,\overline{\epsilon}) \Bigl\{ \left[ G^{-1}(\omega,\epsilon,\overline{\epsilon}) \right]^{\text{R}} \Bigr\}^{-1},
\end{equation}
$\Bigl\{ \left[ G^{-1}(\omega,\epsilon,\overline{\epsilon}) \right]^{\text{R}} \Bigr\}^{-1}$ being the inverse time-averaged retarded component of the GF in Equation~\eqref{eq:FullDysonEq}. 

In terms of the {\em contour-times} $z,z^\prime$, and in the Migdal approximation~\cite{mu.we.15,mu.ts.17,ma.ga.22,ma.we.23}, the e-ph SE reads
\begin{equation}\label{eq:backbone_e-ph_SE}
\Sigma_{\text{e-ph}}(z,z^{\prime}) = \ii g^{2} G_{\text{loc}}(z,z^{\prime}) D_{\text{ph}}(z,z^{\prime})
\end{equation}
and corresponds to the lowest-order expansion in the phonon GF $D_{\text{ph}}$, the form of which will be discussed below. $G_{\text{loc}}(z,z^{\prime})$ is the contour-times local GF, the retarded component of which obeys Equation~\eqref{eq:local_GF}. The {\em retarded} and {\em Keldysh} components of Equation~\eqref{eq:backbone_e-ph_SE} can be found in previous work from the authors~\cite{ma.ga.22}.

In this manuscript we will use the {\em wide-band limit} approximation~\cite{ne.ar.15} for the heat bath described by $\hat{H}_{\text{bath}}$ according to which $\Sigma^{\text{R}}_{\text{bath}}$ reads
\begin{equation}\label{eq:ret_sigma_bath}
 \Sigma^{\text{R}}_{\text{bath}}(\omega) = - \ii \frac{\Gamma_{\text{e}}}{2} ,
\end{equation}
where $\Gamma_{\text{e}}$ is the so-called {\em electronic dephasing rate}~\cite{ma.we.23}. The Keldysh component $\Sigma^{\text{K}}_{\text{bath}}$ is obtained by the {\em fluctuation-dissipation} theorem for fermions, i.e. $\Sigma^{\text{K}}_{\text{bath}}(\omega)= \left[ \Sigma^{\text{R}}_{\text{bath}}(\omega) - \Sigma^{\text{A}}_{\text{bath}}(\omega) \right] \tanh\left[\beta(\omega-\mu)/2\right]$ with $\beta$ the inverse temperature and $\mu$ the chemical potential of the bath. Other important observables are the current $J$ flowing in the direction of the applied field and the kinetic energy $E_{\text{kin}}$ of the electrons of the lattice: the derivation of both these quantities can be found in previous works from the authors~\cite{ma.ga.22,ma.we.23}.

\subsubsection{Phonon Dyson equation}

The optical phonon branch consists of Einstein phonons coupled to an ohmic bath~\cite{mu.ts.17,ma.we.23}, the Dyson equation of which reads
\begin{equation}\label{eq:local_Dyson_Ein_ph}
\kel{D}_{\text{ph}}(\omega) = [\kel{D}^{-1}_{\text{ph},\text{E}}(\omega) - \kel{\Pi}_{\text{bath}}(\omega) - \kel{\Pi}_{\text{e-ph}}(\omega)]^{-1}
\end{equation}
with the non-interacting retarded component of the Einstein phonon GF given by
\begin{equation}\label{eq:non-int_einstein_ph}
D^{\text{R}}_{\text{ph},\text{E}}(\omega) = 2\omega_{\text{E}}/\left(\omega^{2} - \omega_{\text{E}}^{2}\right).
\end{equation}
The Einstein phonon is coupled to an ohmic bath $\hat{H}_{\text{ph},\text{ohm}}$, the real {\em retarded} GF of which is obtained from the Kramers-Kr\"onig relations,~\cite{mu.ts.17} while the {\em Keldysh} component is given by the fluctuation-dissipation theorem for bosons, i.e.
\begin{equation}\label{eq:ohmic_bath_GF}
\Pi^{\text{K}}_{\text{bath}}(\omega) = -2\pi\ii A_{\text{bath}}(\omega) \coth(\beta\omega/2).
\end{equation}
We choose the following form for the ohmic bath density of states (DOS) in~\eqref{eq:ohmic_bath_GF}
\begin{equation}\label{eq:ohm_bath_spec}
A_{\text{bath}}(\omega) = \frac{v^{2}_{\text{c}}}{\omega_{\text{c}}} \left[ \frac{1}{1+\left( \frac{\omega-\omega_{\text{c}}}{\omega_{\text{c}}}\right)^{2}}  - \frac{1}{1+\left( \frac{\omega+\omega_{\text{c}}}{\omega_{\text{c}}}\right)^{2}} \right],
\end{equation}
where $-\pi A_{\text{bath}}(\omega) \equiv \text{Im}[\Pi^{\text{R}}_{\text{bath}}(\omega)]$. In Equation~\eqref{eq:ohm_bath_spec} $\omega_{\text{c}}$ denotes the ohmic bath cutoff frequency and $v_{\text{c}}$ the hybridization strength to the ohmic bath: we only stress that the DOS in Equation~\eqref{eq:ohm_bath_spec} is linear for $\omega \in \left[ -\omega_{\text{c}},\omega_{\text{c}} \right]$.

According to the DMFT approximation, the polarization diagram $\Pi_{\text{e-ph}}$ only depends on the local electron GF. Within the Migdal approximation, the contour times {\em polarization} diagram~\cite{mu.we.15,mu.ts.17} in Equation~\eqref{eq:local_Dyson_Ein_ph} reads
\begin{align}\label{eq:bubble_GG}
\Pi_{\text{e-ph}}(z,z^{\prime})=-2\ii g^{2} G_{\text{loc}}(z,z^{\prime})G_{\text{loc}}(z^{\prime},z),
\end{align}
where the factor $2$ accounts for spin degeneracy. We denote the scheme in which $\Pi_{\text{e-ph}}(z,z^{\prime})$ is set to zero as non-self-consistent (NSC) while within the self-consistent (SC) treatment the phonon SE in Equation~\eqref{eq:bubble_GG} is non-vanishing. The real time components of Equation~\eqref{eq:bubble_GG} have been derived in previous work from the authors~\cite{ma.we.23}.

\subsection{IPT impurity solver and DMFT loop}

The IPT impurity solver is based on the perturbative expansion of the electron SE in terms of the {\em Weiss field}~\cite{ka.ko.96}, namely
\resub{
\begin{equation}\label{eq:weiss_field}
 \mathcal{\kel{G}}^{-1}_{0}(t,t^{\prime}) \equiv \kel{g}^{-1}_{0}(t,t^{\prime}) - \kel{\Delta}(t,t^{\prime}),
\end{equation}
$\kel{g}_{0}$ being the non-interacting GF of the isolated site, the retarded component of which (in frequency domain) reads $g^{\text{R}}_{0}(\omega) = \left( \omega - \varepsilon_{\text{c}} \right)^{-1}$. The quantity $\kel{\Delta}$ in Equation~\eqref{eq:weiss_field} is the {\em hybridization} function of the system, encoding the effects of all the other lattice sites in a mean-field fashion~\cite{me.vo.89,ge.ko.92,ge.ko.96}. The real-time components of the electron SE can then be written as}
\begin{equation}\label{eq:IPT_SEs}
\Sigma^{\lessgtr}(t,t^{\prime}) = U^{2} \mathcal{G}^{\lessgtr}_{0}(t,t^{\prime}) \mathcal{G}^{\gtrless}_{0}(t^{\prime},t) \mathcal{G}^{\lessgtr}_{0}(t,t^{\prime}).
\end{equation}
The {\em retarded} and {\em Keldysh} components of the electron SE can be obtained from Equation~\eqref{eq:IPT_SEs} as
\begin{align}\label{eq:IPT_SEs_RK}
\begin{split}
\Sigma^{\text{R}}(t,t^{\prime}) & = \theta(t-t^{\prime}) \left(\Sigma^{>}(t,t^{\prime}) - \Sigma^{<}(t,t^{\prime}) \right) \\
\Sigma^{\text{K}}(t,t^{\prime}) & = \Sigma^{>}(t,t^{\prime}) + \Sigma^{<}(t,t^{\prime}).
\end{split}
\end{align}
At the steady-state all quantities in Equation~\eqref{eq:IPT_SEs_RK} are dependent on the difference $t-t^{\prime}$ alone, hence their Fourier transform in frequency domain is straightforward. The {\em retarded} and {\em Keldysh} components of the Weiss field then read
\begin{align}\label{eq:IPT_Weiss_field_RK}
\begin{split}
\mathcal{G}^{\text{R}}_{0}(\omega) & = \frac{1}{\Sigma^{\text{R}}(\omega)+G^{-1,\text{R}}_{\text{loc}}(\omega)}, \\
\mathcal{G}^{\text{K}}_{0}(\omega) & = - |\mathcal{G}^{\text{R}}_{0}(\omega)|^{2}\left(\Sigma^{\text{K}}(\omega)-\frac{G^{\text{K}}_{\text{loc}}(\omega)}{|G^{\text{R}}_{\text{loc}}(\omega)|^{2}}\right).
\end{split}
\end{align}
We stress that at half-filling the Hartree term $U/2$ must be explicitly added to $\Sigma^{\text{R}}(\omega)$ before computing the quantities in Equation~(\ref{eq:IPT_Weiss_field_RK}). The main steps of DMFT employing the IPT as impurity solver are:

\begin{itemize}
\item[i.] Guess $\kel{\Sigma}(\omega)$, $\kel{\Sigma}_{\text{e-ph}}(\omega)$ and $\kel{\Pi}_{\text{e-ph}}(\omega)$
\item[ii.] Compute $\kel{G}_{\text{loc}}(\omega)$ and $\kel{D}_{\text{ph}}(\omega)$ as in Equation~\eqref{eq:local_GF} and \eqref{eq:local_Dyson_Ein_ph} 
\item[iii.] Extract $\mathcal{G}^{\lessgtr}_{0}(\omega) = [\mathcal{G}^{\text{K}}_{0}(\omega) \mp (\mathcal{G}^{\text{R}}_{0}(\omega) - \mathcal{G}^{\text{A}}_{0}(\omega) ) ]/2$ from Equation~\eqref{eq:IPT_Weiss_field_RK}
\item[iv.] Fourier-transform $\mathcal{G}^{\lessgtr}_{0}(\omega)$ to get $\Sigma^{\lessgtr}(t,t^{\prime})$ as in~\eqref{eq:IPT_SEs}
\item[v.] Fourier-transform $\kel{G}_{\text{loc}}(\omega)$ and $\kel{D}_{\text{ph}}(\omega)$, compute $\Sigma^{\text{R/K}}_{\text{e-ph}}(t,t^{\prime})$ and $\Pi^{\text{R/K}}_{\text{e-ph}}(t,t^{\prime})$
\item[vi.] Update $\kel{\Sigma}(\omega)$, $\kel{\Sigma}_{\text{e-ph}}(\omega)$ and $\kel{\Pi}_{\text{e-ph}}(\omega)$ via back Fourier-transform.
\end{itemize}

The steps ii. to vi. are then repeated until convergence. 

\section{Results}\label{sec:results}

In this section we benchmark the results yielded by the IPT approach against those obtained within the AMEA impurity solver for the system at hand.

\begin{table}[h]
  \begin{center}
\begin{tabular}{ ccccccccc }
      \hline
      \hline
       $U$ & $\varepsilon_{\text{c}}$ & $\mu$ & $1/\beta$ & $\omega_{\text{c}}$ & $v_{\text{c}}$ & $g$ & $\omega_{\text{E}}$ \\
      \hline
       8 & -4 & 0 & 0.05 & 0.6 & 0.055 & 0.4 & 0.6  \\
      \hline
      \hline
    \end{tabular}
    \caption{Default values of the main parameters used in this manuscript. All values are in units of the hopping $t^{\ast}$.}
    \label{tab:default_pars}
  \end{center}
\end{table}

\subsection{SC and NSC phonons}\label{sec:IPT_Ein_ph_res}

\resub{We set off by analyzing the steady-state current and kinetic energy in both the SC and NSC schemes obtained within the IPT impurity solver. We then compare these results with those obtained within the AMEA only at a later time.}

\subsubsection{Steady-state current and kinetic energy}\label{sec:NESS_current}

\begin{figure}[h]
\includegraphics[width=\linewidth]{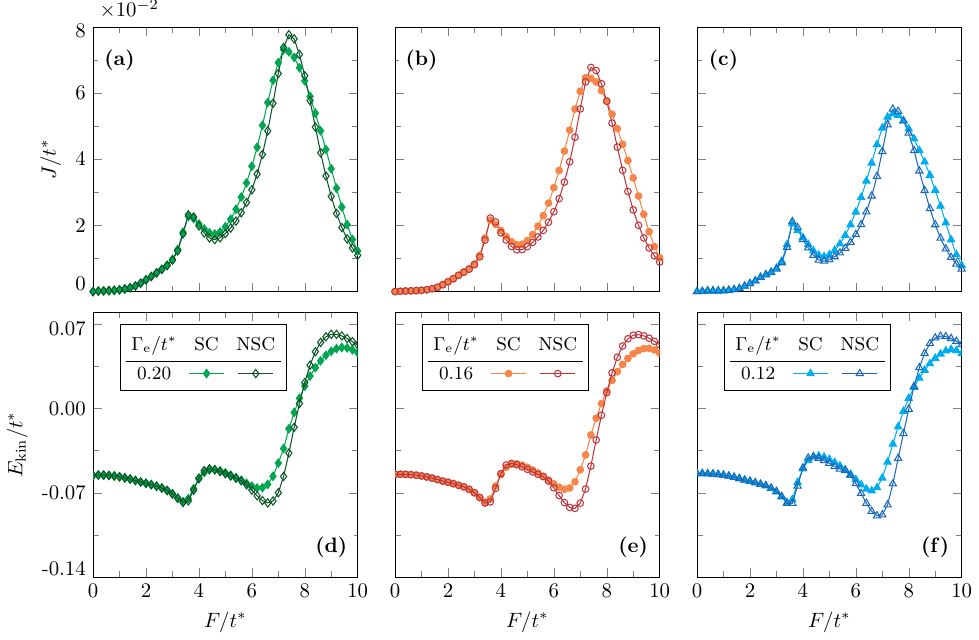}
\caption{Current $J$ as function of the applied field $F$ for (a) $\Gamma_{\text{e}}=0.20t^{\ast}$ (b) $\Gamma_{\text{e}}=0.16t^{\ast}$ and (c) $\Gamma_{\text{e}}=0.12t^{\ast}$ for both SC and NSC schemes. (d), (e) and (f) show the kinetic energy $E_{\text{kin}}$ as function of $F$ for the same coupling strengths, respectively. Default parameters are specified in Table~\ref{tab:default_pars}. \resub{Results have been obtained using the IPT impurity solver.} (Here $U=8t^{\ast}$.)}
\label{fig:Curr_KinEn_IPT}
\end{figure}

The current $J$ and and kinetic energy $E_{\text{kin}}$ as functions of the applied field $F$ for both the SC and NSC schemes obtained employing the IPT impurity solver are shown in \textbf{Figure~\ref{fig:Curr_KinEn_IPT}}.
We observe that the SC treatment has no effect for field strengths $F<U/2$ as the SC and NSC curves for both $J$ and $E_{\text{kin}}$ lie on top of each other. On the other hand, the SC scheme does contribute corrections, albeit tiny, once $F$ gets past field strengths of the order of the resonance $U/2$, for both $J$ and $E_{\text{kin}}$. In particular \resub{we observe a suppression of the peak values of the current $J$ in the SC approach at $F\approx U$ with respect to the NSC treatment. This is accompanied by an overall smearing of the $J$-$F$ curves around the maximum for all the values of $\Gamma_{\text{e}}$ employed in this manuscript, see Figure~\ref{fig:Curr_KinEn_IPT}(a),~\ref{fig:Curr_KinEn_IPT}(b) and~\ref{fig:Curr_KinEn_IPT}(c)}. \resub{The kinetic energy is also affected by the SC treatment as its maximum value is suppressed, while its minimum is raised with respect to the NSC scheme regardless of the value of $\Gamma_{\text{e}}$, see Figure~\ref{fig:Curr_KinEn_IPT}(d),~\ref{fig:Curr_KinEn_IPT}(e) and~\ref{fig:Curr_KinEn_IPT}(f).}

\begin{figure}[h]
\includegraphics[width=\linewidth]{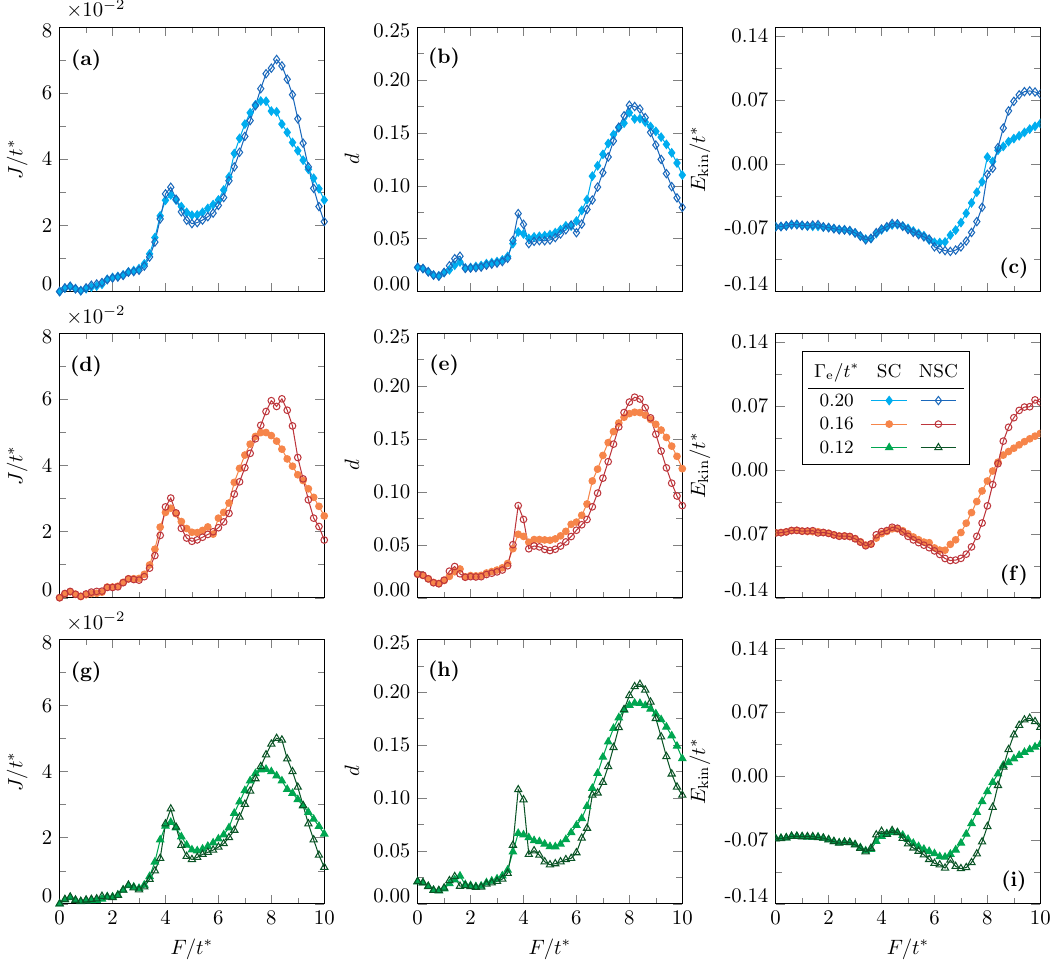}
\caption{(a) Current, (b) double occupation and (c) kinetic energy as function of the applied field $F$ at $\Gamma_{\text{e}}=0.20t^{\ast}$ for both SC and NSC schemes. (d), (e) and (f) show the same quantities for $\Gamma_{\text{e}}=0.16t^{\ast}$, while (g), (h) and (i) refer to $\Gamma_{\text{e}}=0.12t^{\ast}$. Default parameters are specified in Table~\ref{tab:default_pars}. \resub{These results have been obtained using the AMEA impurity solver and originally published by the authors in \url{https://link.aps.org/doi/10.1103/PhysRevB.107.155103}.} (Here $U=8t^{\ast}$.)}
\label{fig:fig1_prb}
\end{figure}

This behavior is qualitatively in agreement with the results \resub{obtained within the AMEA impurity solver presented in a recent work from the authors}~\cite{ma.we.23}. \resub{As a matter of fact, using AMEA the drop in the current $J$ at the resonance $F\approx U$ observed in the SC scheme is way more pronounced than in the IPT case as one can see by comparing Figure~\ref{fig:Curr_KinEn_IPT} and~\ref{fig:fig1_prb} for corresponding values of the dephasing rate $\Gamma_{\text{e}}$. On the other hand, the differences between the impurity solvers are less pronounced when it comes to the kinetic energy $E_{\text{kin}}$, as it can be observed by comparing the corresponding curves in Figure~\ref{fig:Curr_KinEn_IPT} and~\ref{fig:fig1_prb} for the same $\Gamma_{\text{e}}$'s.} 

We want to mention that the orders of magnitude of both $J$ and $E_{\text{kin}}$ in the IPT scheme quantitatively agrees with those obtained within the AMEA impurity solver. However, using the IPT approach the two main resonances at $F\approx U/2$ and $F\approx U$ are shifted towards smaller field strengths and the differences between the SC and NSC schemes are not as pronounced as in the AMEA scheme [see again Figure~\ref{fig:fig1_prb}]. 

\resub{In addition, the two impurity solvers differ in that the IPT cannot capture correlation effects like the {\em tiny} resonance in the current $J$ at $F\sim U/3 \approx 2.6$~\footnote{For further details about the resonances in the $J$-$F$ curves we refer to the previous work~\cite{mu.we.18,ma.ga.22}.}, which instead can be distinguished using the AMEA solver as soon as $\Gamma_{\text{e}}$ is small enough, see Figure~\ref{fig:fig1_prb}(a),~\ref{fig:fig1_prb}(d) and in particular~\ref{fig:fig1_prb}(g).}
\begin{figure}[t]
\includegraphics[width=\linewidth]{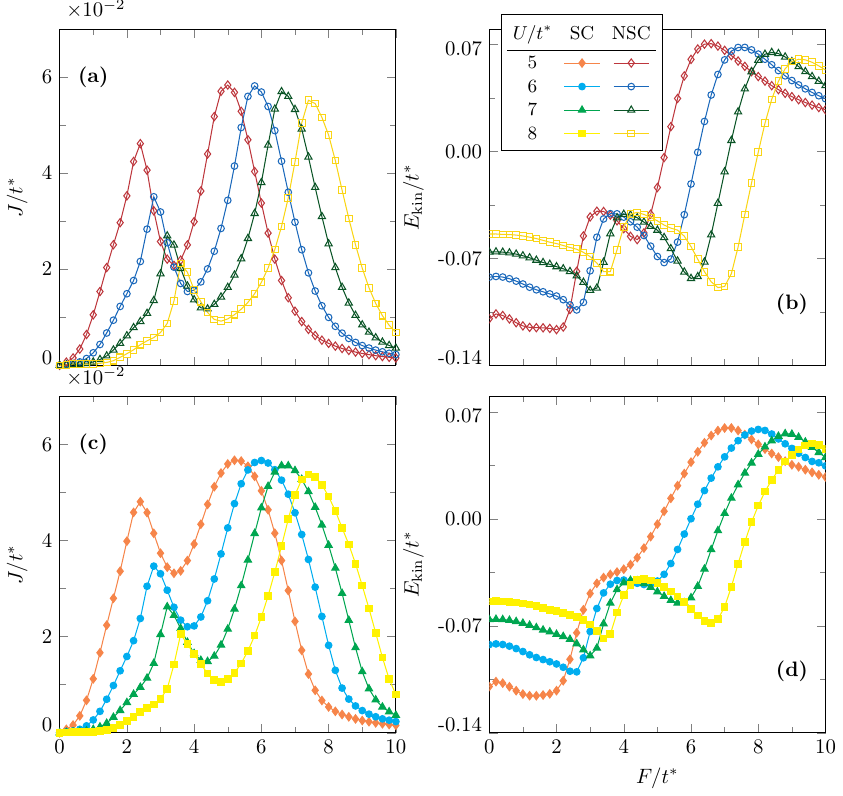}
\caption{Current $J$ within the (a) NSC and (c) SC scheme for selected values of the Hubbard $U$ as function of the applied field $F$. (b) and (d) show the kinetic energy $E_{\text{kin}}$ as function of $F$ in the NSC and SC scheme, respectively, for the same values of the Hubbard $U$. Default parameters are specified in Table~\ref{tab:default_pars}. \resub{Results have been obtained using the IPT impurity solver.} (Here $\Gamma_{\text{e}}=0.12t^{\ast}$.)}
\label{fig:Curr_KinEn_Us_IPT}
\end{figure}

\begin{figure}[t]
\includegraphics[width=\linewidth]{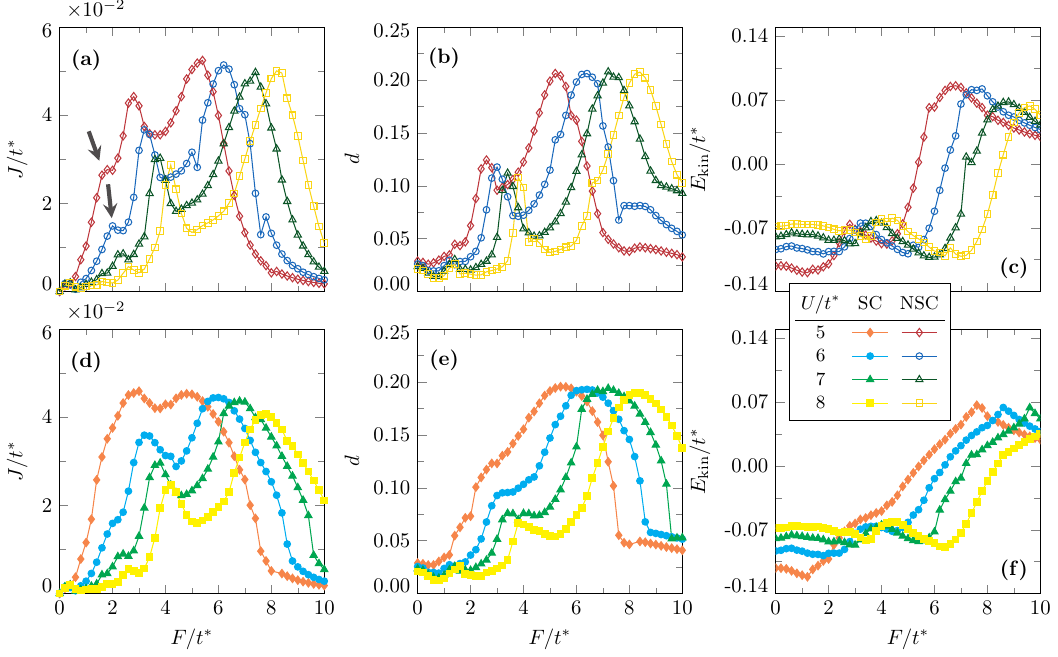}
\caption{Current $J$ within (a) NSC and (d) SC scheme for selected values of the Hubbard $U$ as function of the applied field $F$. Black arrows in (a) highlight the progressive {\em merge} of the resonances at $F\sim U/3$ and $F=U/2$ as $U$ is lowered which is enhanced by the SC treatment. (b) and (e) show the double occupation $d$ as function of $F$ for the NSC and SC scheme, respectively, while the kinetic energy $E_{\text{kin}}$ is shown in (c) and (f) for the same values of $U$. Default parameters are specified in Table~\ref{tab:default_pars}. \resub{These results have been obtained using the AMEA impurity solver and originally published by the authors in \url{https://link.aps.org/doi/10.1103/PhysRevB.107.155103}.} (Here $\Gamma_{\text{e}}=0.12t^{\ast}$.)}
\label{fig:fig5_prb}
\end{figure}

\subsubsection{The effect of the Hubbard $U$}

\resub{This section is devoted to the analysis of the role of the Hubbard $U$ on the insulating phase: once again we focus on the current $J$ and kinetic energy $E_{\text{kin}}$ in the SC and NSC approaches to then compare them to the corresponding quantities obtained within the AMEA impurity solver.}

In \textbf{Figure~\ref{fig:Curr_KinEn_Us_IPT}} is displayed the behavior of $J$ and $E_{\text{kin}}$ as function of the applied field for selected values of $U$, see panels (a) and (b) for the NSC scheme and (c) and (d) for SC treatment. We observe that in the SC case both the $J$-$F$ and $E_{\text{kin}}$-$F$ curves are broadened, see panels (b) and (d).

\resub{In particular we observe an increase in the current $J$ at $F\approx U/2$ as $U$ is decreased for both NSC and SC schemes. Also, while the values of the two peaks at $U/2$ and $U$ stay approximately the same in height in the NSC and SC schemes, we find that the latter treatment enhances $J$ for field strengths that lie in between the two main resonances, compare Figure~\ref{fig:Curr_KinEn_Us_IPT}(a) and~\ref{fig:Curr_KinEn_Us_IPT}(c).}

\resub{When it comes to the kinetic energy, on the other hand, let alone an overall {\em smearing} of the curves the SC treatment does not affect $E_{\text{kin}}$ significantly, compare Figure~\ref{fig:Curr_KinEn_Us_IPT}(b) and~\ref{fig:Curr_KinEn_Us_IPT}(d). As already discussed in Reference~\cite{aron.12} both the $J$-$F$ and $E_{\text{kin}}$-$F$ curves collapse on one another when they are plotted as function of the difference $F-U$ (not shown), signalling that the breakdown of the insulating phase depends on the value of the Hubbard $U$ alone.}

Once again we stress the qualitative agreement between the results of Figure~\ref{fig:Curr_KinEn_Us_IPT} and those presented in the previous work~\cite{ma.we.23} obtained within AMEA that we reproduce in Figure~\ref{fig:fig5_prb}. \resub{However, the two solvers do differ in that, as it can be shown with the help of Figure~\ref{fig:Curr_KinEn_IPT}(c) and~\ref{fig:fig1_prb}(g), the IPT approach fails to capture the resonance in $J$ at $F\sim U/3$~\footnote{We stress that this resonance can be distinguished only when the electron dephasing rate $\Gamma_{\text{e}}$ is small. For further details about the role of $\Gamma_{\text{e}}$ in dissipative Mott insulating systems we refer to our recent work~\cite{ma.ga.22,ma.we.24}.}.}

\resub{We see that such resonance is missing within the IPT solver in both the SC and NSC schemes regardless of the value of $U$, see Figure~\ref{fig:Curr_KinEn_Us_IPT}(a) and~\ref{fig:Curr_KinEn_Us_IPT}(c). On the other hand, when using the AMEA one can appreciate it, even though it tends to be smeared out by the SC treatment, especially for small values of $U$, see Figure~\ref{fig:fig5_prb}(a) and~\ref{fig:fig5_prb}(d).}

\resub{Furthermore, the enhancement of the current $J$ in between the two main resonances $F=U/2$ and $U$ occurring in the SC treatment within the IPT approach [see again Figure~\ref{fig:Curr_KinEn_Us_IPT}(a) and~\ref{fig:Curr_KinEn_Us_IPT}(c)] is even more pronounced when the AMEA impurity solver is used, especially for small values of $U$, as one can see by comparing the curves for the current $J$ in Figure~\ref{fig:fig5_prb}(a) and~\ref{fig:fig5_prb}(d).}

\resub{The main difference in the kinetic energy between the ITP and AMEA impurity solvers lies in the overall {\em sharper} $E_{\text{kin}}$-$F$ curves in the latter approach, as one can see by comparing the results shown in Figure~\ref{fig:Curr_KinEn_Us_IPT} and~\ref{fig:fig5_prb}. It is worth mentioning that the SC treatment smears out the $E_{\text{kin}}$-$F$ curves more when the AMEA solver is employed especially for field strengths $F\sim U/2$.}

\subsubsection{Equilibrium spectral features}

\resub{In this section we briefly compare the electronic spectral features at equilibrium, i.e. $F=0$.}

\begin{figure}[b]
\includegraphics[width=\linewidth]{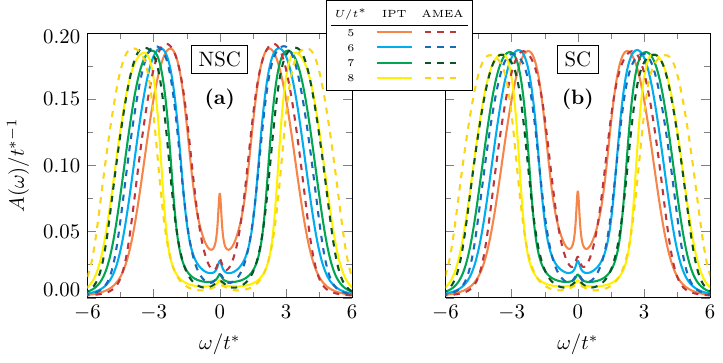}
\caption{Electron SF $A(\omega)$ at equilibrium ($F=0$) for the (a) NSC and (b) SC scheme as function of selected values of the Hubbard $U$ obtained within the IPT and AMEA impurity solvers. Default parameters are specified in Table~\ref{tab:default_pars}. (Here $\Gamma_{\text{e}}=0.12t^{\ast}$.)}
\label{fig:QPP_SFs_IPT_vs_AMEA}
\end{figure}

\textbf{Figure~\ref{fig:QPP_SFs_IPT_vs_AMEA}} shows the electron SF at $F=0$ within the (a) NSC and (b) SC schemes for both the IPT and AMEA impurity solvers. The IPT-resolved SF shows a much more pronounced {\em quasi-particle} peak (QPP) at around $\omega=0$~\cite{ma.ga.22} as $U$ is reduced, together with an underestimation of the width of the Hubbard bands within both the NSC and SC schemes, as compared to the results obtained with the AMEA impurity solver. On the other hand, at $U=8t^{\ast}$ the height of the QPP in the IPT and AMEA approaches is in quite good agreement, even though the Hubbard bands are still narrower in the IPT scheme in both the NSC and SC approaches. 

\resub{The larger amount of {\em in-gap states} due to the QPP at $\omega \approx 0$ in the IPT with respect to AMEA is common to all electric field strengths used in this manuscript (not shown) and reveals the systematic underestimation of the band gap committed by the IPT solver. This underestimation, in turn, explains the {\em shifting} of the main resonances (at $F=U/2$ and $U$) in the current $J$ toward smaller values of the electric field discussed in Section~\ref{sec:NESS_current}.}

\subsection{The threshold field}

\begin{figure}[b]
\includegraphics[width=\linewidth]{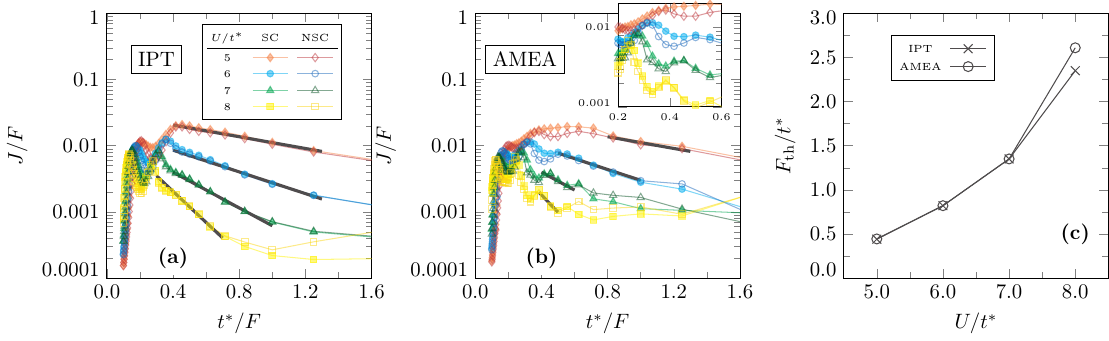}
\caption{(a) Ratio $J/F$ as function of the inverse field $1/F$ for selected values of the Hubbard $U$ within the NSC and SC schemes obtained with IPT impurity solver. \resub{(b) shows the same quantity obtained within the AMEA: it is worth noticing the kink at $F^{-1}\sim 0.4t^{\ast -1}$ (see inset) corresponding to the resonance at $F\approx U/3$ which is absent in the IPT scheme, see panel (a). Black straight lines in both (a) and (b) extend over the range of values of $1/F$ used for the linear regression fit yielding the threshold field $F_{\text{th}}$ as function of $U$ shown in (c) for the two impurity solvers.} The SC treatment has no effect on $F_{\text{th}}$, which is the reason why panel (c) does not distinguish between the two cases. Default parameters are specified in Table~\ref{tab:default_pars}. (Here $\Gamma_{\text{e}}=0.12t^{\ast}$.)}
\label{fig:Curr_threshold_Us_IPT}
\end{figure}

\resub{In models of Mott insulating systems one expects the $J$-$F$ curve to display the {\em threshold} behavior~\cite{ec.ok.10,ec.we.13.db}
\begin{equation}\label{eq:threshold_field_J}
J\propto F \exp(-F_{\text{th}}/F),
\end{equation}
due to the opening of a gap that prevents the free motion of electrons from the lower to the upper band. Equation~\eqref{eq:threshold_field_J} shows that only when the applied field $F$ gets past the threshold $F_{\text{th}}$ can electrons cross the band gap and thus give rise to a steady-state current. By determining the threshold $F_{\text{th}}$ one can then infer the magnitude of the {\em effective}~\footnote{Here {\em effective} is opposed to nominal gap, i.e. the value of the Hubbard interaction $U$.} band gap of the model. 
However, as it has been shown in previous works investigating Mott insulators with a large~\footnote{The gap must exceed the characteristic energy scale provided by the hopping $t^{\ast}$ by several times at least.} band gap~\cite{mu.we.18,ma.ga.22}, correlation effects are responsible for resonances in the current $J$ at $F\approx U/n$, see also Section~\ref{sec:IPT_Ein_ph_res} [Figure~\ref{fig:Curr_KinEn_IPT} and~\ref{fig:fig1_prb}], so we may expect the existence of at least two threshold fields.

It is worth recalling that these resonances are determined by emergence of the {\em Wannier-Stark}~\cite{mu.we.18,ma.ga.22} states in the local electron SF which effectively allow electron tunnelling to the upper band by {\em bridging} the band gap and are sometimes referred to as {\em Landau-Zener} excitations in other models~\cite{aron.12}.

However, in this manuscript, $F_{\text{th}}$ identifies the field strength necessary for the current $J$ to reach non-negligible values for the {\em first} time, so Equation~\eqref{eq:threshold_field_J} should hold for $F<U/2$ at least for the IPT solver as it cannot capture the whole of electronic correlation~\footnote{This is due to the fact that the electron SE is constructed from a noninteracting GF, the Weiss field in Equation~\eqref{eq:IPT_SEs}.}, i.e. there is no other resonance before field strengths of the order of half of the band gap.

This section is then devoted to the analysis of these aspects limited to the comparison of the IPT and AMEA solvers. In fact, given the simple model at hand the study of the dependence of the effective band gap on the threshold $F_{\text{th}}$ goes beyond the purpose of this manuscript: for a detailed study can be found in the previous work~\cite{ok.ao.10,ec.ok.10}.}

We benchmark the results for the $J$-$F$ curve obtained within the IPT and AMEA impurity solvers for selected values of $U$ against a linear regression fit according to Equation~\eqref{eq:threshold_field_J}. \textbf{Figure~\ref{fig:Curr_threshold_Us_IPT}}(a) shows the ratio $J/F$ as function of $1/F$ in both SC and NSC schemes \resub{obtained using the IPT}: the values of the inverse field used for the fit can be deduced by the extent of the black line. \resub{As one can see by direct inspection the curves for all values of $U$ exhibit a linear behavior for a wide range of values of inverse field strengths up to $F^{-1}\sim 0.25t^{\ast -1}$, which correspond to the resonance at $F=U/2$.}

\resub{The results obtained with the AMEA impurity solver, instead, are shown in Figure~\ref{fig:Curr_threshold_Us_IPT}(b), see corresponding inset for a close-up of the region around $F^{-1}\sim 0.4t^{\ast -1}$, corresponding to $F\approx U/3$. The most evident feature is the small {\em kink} occurring at $F^{-1} \sim 0.4t^{\ast -1}$, clearly visible especially for $U>5t^{\ast}$, which is absent in the IPT approach as it can be seen by comparing Figure~\ref{fig:Curr_threshold_Us_IPT}(a) and~\ref{fig:Curr_threshold_Us_IPT}(b). Due to this additional resonance, the linear regression fit within the AMEA impurity solver can be performed over a far smaller range of values of $F^{-1}$ and indeed there are two regions where linearity holds: the first occurs for $F^{-1}<0.4t^{\ast -1}$ and the second for inverse field smaller than $0.25t^{\ast -1}$, see again Figure~\ref{fig:Curr_threshold_Us_IPT}(b).} 

The threshold fields obtained with the linear regression fit are shown in Figure~\ref{fig:Curr_threshold_Us_IPT}(c): the SC treatment basically does not affect the results within the numerical accuracy, thus leaving the threshold fields unaltered in both the IPT and AMEA approaches. \resub{It should be noted that by performing the linear regression fit for $F^{-1}<0.4t^{\ast -1}$ as in this case, the IPT- and AMEA-resolved curves are basically on top of one another. The AMEA scheme, however, yields a slightly larger threshold $F_{\text{th}}$ than the IPT one for $U=8t^{\ast}$, see again Figure~\ref{fig:Curr_threshold_Us_IPT}(c).} As expected, a larger $F_{\text{th}}$ is required to compensate for a larger band gap (and hence a larger $U$) and promote particles across it.  \resub{However, due to the extension of the Hubbard bands the effective gap, and thus the threshold field, is way smaller than the Hubbard $U$ so that a naive relation of the form $F_{\text{th}} \sim U$ does not hold (see Figure~\ref{fig:Curr_threshold_Us_IPT}(c)), as already argued in previous work~\cite{ec.ok.10}}.

\section{Conclusion}\label{sec:conclusion}

In this manuscript a Mott insulating system has been characterized in terms of its conducting properties when subject to an external static electric. Optical phonons and electronic heat bath provide the relaxation pathways for the extra energy injected by the field, so that the electron of the lattice can relax back to the valence band and a steady-state current be established. 

The iterated perturbation theory (IPT) approach has been used as impurity solver to address the steady-state of the system. The corresponding results have been benchmarked against those obtained within a much more computationally costly impurity solver developed by the authors, the so-called auxiliary master equation approach (AMEA). It has been shown that the results obtained employing the IPT qualitatively agree with those of the AMEA impurity solver even though the former approach does not capture the resonances in the current characteristics which are directly related to the correlated nature of the electrons of the lattice. Being computationally cheaper, the IPT solver could be used to span the parameter space when investigating novel setups and gather information about the interesting regions to be addressed by a more reliable and computationally costly impurity solver.

\medskip
\textbf{Acknowledgements} \par 
This research was funded by the Austrian Science Fund (Grant No. P 33165-N) and by NaWi Graz. The results have been obtained using the Vienna Scientific Cluster and the D-Cluster Graz.


\begin{figure}
\textbf{Table of Contents}\\
\medskip
  \includegraphics[height=50mm,width=55mm]{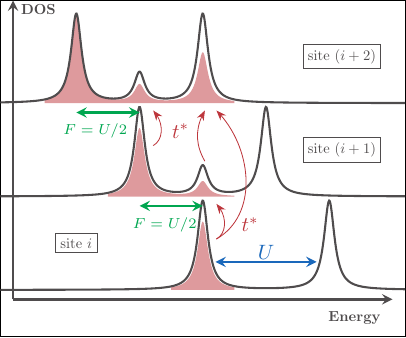}
  \medskip
  \caption*{Schematic representation of the current flowing within the lattice. When the field is applied electrons can tunnel through the former gap of an adjacent site, creating the necessary states to populate the conduction band of a site which is twice lattice spacings apart.}
\end{figure}

\bibliographystyle{MSP}
\bibliography{references_database,my_refs}

\begin{thebibliography}{10}
\providecommand{\url}[1]{\texttt{#1}}
\providecommand{\urlprefix}{URL }

\bibitem{aron.12}
C.~Aron,
\newblock \emph{Phys. Rev. B} \textbf{2012}, \emph{86} 085127.

\bibitem{am.we.12}
A.~Amaricci, C.~Weber, M.~Capone, G.~Kotliar,
\newblock \emph{Phys. Rev. B} \textbf{2012}, \emph{86} 085110.

\bibitem{ar.ko.12}
C.~Aron, G.~Kotliar, C.~Weber,
\newblock \emph{Phys. Rev. Lett.} \textbf{2012}, \emph{108} 086401.

\bibitem{li.ar.15}
J.~Li, C.~Aron, G.~Kotliar, J.~E. Han,
\newblock \emph{Phys. Rev. Lett.} \textbf{2015}, \emph{114} 226403.

\bibitem{ha.li.18}
J.~E. Han, J.~Li, C.~Aron, G.~Kotliar,
\newblock \emph{Phys. Rev. B} \textbf{2018}, \emph{98} 035145.

\bibitem{mu.we.18}
Y.~Murakami, P.~Werner,
\newblock \emph{Phys. Rev. B} \textbf{2018}, \emph{98} 075102.

\bibitem{le.ra.85}
P.~A. Lee, T.~V. Ramakrishnan,
\newblock \emph{Rev. Mod. Phys.} \textbf{1985}, \emph{57} 287.

\bibitem{be.ki.94}
D.~Belitz, T.~R. Kirkpatrick,
\newblock \emph{Rev. Mod. Phys.} \textbf{1994}, \emph{66} 261.

\bibitem{st.ca.13}
P.~Stoliar, L.~Cario, E.~Janod, B.~Corraze, C.~Guillot-Deudon,
  S.~Salmon-Bourmand, V.~Guiot, J.~Tranchant, M.~Rozenberg,
\newblock \emph{Advanced Materials} \textbf{2013}, \emph{25}, 23 3222.

\bibitem{ja.tr.15}
E.~Janod, J.~Tranchant, B.~Corraze, M.~Querr{\'e}, P.~Stoliar, M.~Rozenberg,
  T.~Cren, D.~Roditchev, V.~T. Phuoc, M.-P. Besland, L.~Cario,
\newblock \emph{Advanced Functional Materials} \textbf{2015}, \emph{25}, 40
  6287.

\bibitem{me.vo.89}
W.~Metzner, D.~Vollhardt,
\newblock \emph{Phys. Rev. Lett.} \textbf{1989}, \emph{62} 324.

\bibitem{ge.ko.92}
A.~Georges, G.~Kotliar,
\newblock \emph{Phys. Rev. B} \textbf{1992}, \emph{45} 6479.

\bibitem{ge.ko.96}
A.~Georges, G.~Kotliar, W.~Krauth, M.~J. Rozenberg,
\newblock \emph{Rev. Mod. Phys.} \textbf{1996}, \emph{68} 13.

\bibitem{ko.sa.06}
G.~Kotliar, S.~Y. Savrasov, K.~Haule, V.~S. Oudovenko, O.~Parcollet, C.~A.
  Marianetti,
\newblock \emph{Rev. Mod. Phys.} \textbf{2006}, \emph{78}, 3 865.

\bibitem{fr.tu.06}
J.~K. Freericks, V.~M. Turkowski, V.~Zlati{\'{c}},
\newblock \emph{Phys. Rev. Lett.} \textbf{2006}, \emph{97}, 26 266408.

\bibitem{ao.ts.14}
H.~Aoki, N.~Tsuji, M.~Eckstein, M.~Kollar, T.~Oka, P.~Werner,
\newblock \emph{Rev. Mod. Phys.} \textbf{2014}, \emph{86} 779.

\bibitem{ka.ko.96}
H.~Kajueter, G.~Kotliar,
\newblock \emph{Phys. Rev. Lett.} \textbf{1996}, \emph{77} 131.

\bibitem{ar.kn.13}
E.~Arrigoni, M.~Knap, W.~von~der Linden,
\newblock \emph{Phys. Rev. Lett.} \textbf{2013}, \emph{110} 086403.

\bibitem{do.nu.14}
A.~Dorda, M.~Nuss, W.~von~der Linden, E.~Arrigoni,
\newblock \emph{Phys. Rev. B} \textbf{2014}, \emph{89} 165105.

\bibitem{do.ga.15}
A.~Dorda, M.~Ganahl, H.~G. Evertz, W.~von~der Linden, E.~Arrigoni,
\newblock \emph{Phys. Rev. B} \textbf{2015}, \emph{92} 125145.

\bibitem{ti.do.15}
I.~Titvinidze, A.~Dorda, W.~von~der Linden, E.~Arrigoni,
\newblock \emph{Phys. Rev. B} \textbf{2015}, \emph{92} 245125.

\bibitem{we.lo.23}
D.~Werner, J.~Lotze, E.~Arrigoni,
\newblock \emph{Phys. Rev. B} \textbf{2023}, \emph{107} 075119.

\bibitem{ma.we.23}
T.~M. Mazzocchi, D.~Werner, P.~Gazzaneo, E.~Arrigoni,
\newblock \emph{Phys. Rev. B} \textbf{2023}, \emph{107} 155103.

\bibitem{tu.fr.06}
V.~M. Turkowski, J.~K. Freericks,
\newblock \emph{Phys. Rev. B} \textbf{2006}, \emph{73}, 7 075108.

\bibitem{ts.ok.08}
N.~Tsuji, T.~Oka, H.~Aoki,
\newblock \emph{Phys. Rev. B} \textbf{2008}, \emph{78}, 23 235124.

\bibitem{ma.ga.22}
T.~M. Mazzocchi, P.~Gazzaneo, J.~Lotze, E.~Arrigoni,
\newblock \emph{Phys. Rev. B} \textbf{2022}, \emph{106} 125123.

\bibitem{schw.61}
J.~Schwinger,
\newblock \emph{J. Math. Phys.} \textbf{1961}, \emph{2} 407.

\bibitem{keld.65}
L.~V. Keldysh,
\newblock \emph{Sov. Phys. JETP} \textbf{1965}, \emph{20} 1018.

\bibitem{ra.sm.86}
J.~Rammer, H.~Smith,
\newblock \emph{Rev. Mod. Phys.} \textbf{1986}, \emph{58} 323.

\bibitem{ha.ja}
H.~Haug, A.-P. Jauho,
\newblock \emph{Quantum Kinetics in Transport and Optics of Semiconductors},
\newblock Springer, Heidelberg, \textbf{1998}.

\bibitem{st.va.13}
G.~Stefanucci, R.~van Leeuwen,
\newblock \emph{Nonequilibrium Many-Body Theory of Quantum Systems: A modern
  introduction},
\newblock Cambridge University Press, Cambridge, \textbf{2013}.

\bibitem{ec.we.13.db}
M.~Eckstein, P.~Werner,
\newblock \emph{J. Phys.: Conf. Series} \textbf{2013}, \emph{427} 012005.

\bibitem{di.ha.23}
M.~I. D\'{\i}az, J.~E. Han, C.~Aron,
\newblock \emph{Phys. Rev. B} \textbf{2023}, \emph{107} 195148.

\bibitem{ha.ar.23}
J.~E. Han, C.~Aron, X.~Chen, I.~Mansaray, J.-H. Han, K.-S. Kim, M.~Randle,
  J.~P. Bird,
\newblock \emph{Nature Communications} \textbf{2023}, \emph{14}, 1.

\bibitem{mu.we.15}
Y.~Murakami, P.~Werner, N.~Tsuji, H.~Aoki,
\newblock \emph{Phys. Rev. B} \textbf{2015}, \emph{91} 045128.

\bibitem{mu.ts.17}
Y.~Murakami, N.~Tsuji, M.~Eckstein, P.~Werner,
\newblock \emph{Phys. Rev. B} \textbf{2017}, \emph{96} 045125.

\bibitem{jo.fr.08}
A.~V. Joura, J.~K. Freericks, T.~Pruschke,
\newblock \emph{Phys. Rev. Lett.} \textbf{2008}, \emph{101}, 19 196401.

\bibitem{so.do.18}
M.~E. Sorantin, A.~Dorda, K.~Held, E.~Arrigoni,
\newblock \emph{Phys. Rev. B} \textbf{2018}, \emph{97} 115113.

\bibitem{ga.ma.22}
P.~Gazzaneo, T.~M. Mazzocchi, J.~Lotze, E.~Arrigoni,
\newblock \emph{Phys. Rev. B} \textbf{2022}, \emph{106} 195140.

\bibitem{ne.ar.15}
J.~Neumayer, E.~Arrigoni, M.~Aichhorn, W.~von~der Linden,
\newblock \emph{Phys. Rev. B} \textbf{2015}, \emph{92} 125149.

\end{thebibliography}

\end{document}